# Physics-oriented learning of nonlinear Schrödinger equation: optical fiber loss and dispersion profile identification

Takeo Sasai, Masanori Nakamura, Etsushi Yamazaki, Shuto Yamamoto, Hideki Nishizawa, and Yoshiaki Kisaka

Affiliations

NTT Network Innovation Laboratories

Contributions

T. S. conducted all the experiments. T. S. and M. N. developed the code for loss and dispersion monitoring and the signal demodulation process. M. N., E. Y., and S. Y. provided essential support and advices for experimental setup and interpretation of obtained results. H. N and Y. K. guided the method's possible applications. T.S. wrote the paper, and all the co-authors reviewed it and provided comments.

Corresponding author

Takeo Sasai

**Abstract**

In optical fiber communication, system identification (SI) for the nonlinear Schrödinger equation (NLSE) has long been studied mainly for fiber nonlinearity compensation (NLC). One recent line of inquiry to combine a behavioral-model approach like digital backpropagation (DBP) and a data-driven approach like neural network (NN). These works are aimed for more NLC gain; however, by directing our attention to the learned parameters in such a SI process, system status information, i.e., optical fiber parameters, will possibly be extracted. Here, we show that the model-based optimization and interpretable nature of the learned parameters in NN-based DBP enable transmission line monitoring, fully extracting the actual in-line NLSE parameter distributions. Specifically, we demonstrate that longitudinal loss and dispersion profiles along a multi-span link can be obtained at once, directly from data-carrying signals without any dedicated analog devices such as optical time-domain reflectometry. We apply the method to a long-haul (~2,080 km) link and various link conditions are tested, including excess loss inserted, different fiber input power, and non-uniform level diagram. The measurement performance is also investigated in terms of measurement range, accuracy, and fiber launch power. These results provide a path toward simplified and automated network management as another application of DBP.

## Introduction

System identification (SI) for partial differential equations (PDEs), which cannot be analytically solved, has long been studied[1,2]. The task is to find a simplified model or identify system parameters in PDEs from the inputs and outputs of the system. By doing so, the following applications become possible:

(i) the system outputs for arbitrary inputs (and vice versa) become predictable, often with a lower computational cost.

(ii) identification of the parameters in PDEs can reveal the system's status and enable monitoring of it.

For application (i), in the case of optical communication, simplification of the nonlinear Schrödinger equation (NLSE), which governs the optical signal propagation in an optical fiber, has intensively studied since it provides equalization methods for chromatic dispersion (CD) and nonlinear phase rotation (NLPR) with lower computational complexity and facilitates increased data transmission capacity. One of the most common algorithms to compensate the interplay of CD and NLPR via digital signal processing (DSP) is called digital backpropagation (DBP)[3], where signals are backpropagated in virtual links by applying a numerical analysis method for the NLSE, the split-step Fourier method (SSFM) in the reverse direction. However, DBP requires massive computation and thus, various simplified models for the NLSE have been proposed[4,5].

In optical communication, application (ii), which is the target of this study, corresponds to identifying the NLSE parameter distributions from transmitted and received signals. Such distributions enable monitoring of the loss and CD profiles along optical fibers, which are essential measurements before building optical networks and when deducing the causes of network failures. Conventionally, the loss and dispersion have been measured with analog optical devices such as optical time-domain reflectometry (OTDR) and dispersion analyzers, which require injection of optical testing probes into fibers. Such analog approaches are costly, especially in multi-span transmission links, because they require on-site measurements of every optical fiber. If these profiles can be obtained via a digital SI approach, we can expect "smart optical transmission systems," in which all necessary measurement, network building, and fault detection can be conducted automatically without system reconfiguration[6]. Accordingly, signal power (loss) profile extraction of a 260-km multi-span link only from the received signals has been successfully demonstrated[7].

To obtain the benefits of the two applications described above, there are two main approaches for SI: the black-box type and the behavioral-model type. As an example of the former, machine learning methods have gained much attention in the last decade for their potential to model in-out nonlinear dynamics[8]. Though these methods are powerful tools, they tend to be black boxes, which makes it difficult to reveal internal physical characteristics and interpret learned parameters. Thus, a recent line of inquiry is to combine the data-driven approach and more structure-aware models. Interestingly, in the case of optical fiber communication, the well-known SSFM for the NLSE has a structure that is quite similar to or essentially the same as that of neural networks (NNs), in the sense that they both entail a cascaded iteration of linear and nonlinear operations, as illustrated in Fig. 1. With inspiration from this similarity, efficient fiber nonlinearity compensation (NLC) was proposed by regarding DBP as a deep NN[9,10], and the approach has been experimentally verified[11]. Furthermore, by exploiting the interpretable nature of NN-based DBP, a semi-theoretical understanding of the learned CD filters and nonlinear parameters has also been provided[12]. Although these studies showed a significant compensation gain and indicated a new direction for NLC, they fall under the scope of application (i) above, and they only focused on NLC. Consequently, it has not been confirmed whether the learned parameters actually reflect the transmission line status in detail, including characteristics such as the fiber loss, fiber launch power, and fiber types. However, if we turn our attention to the learned parameters rather than concentrate on the NN's output and its accuracy, it might provide a path to application (ii), namely, system status monitoring.

Here, we show that incorporating NN into DBP enables transmission line monitoring and such physics-oriented learning can even identify anomaly losses, fiber launch power, CD, and their positions. Specifically, the longitudinal loss and dispersion (NLSE parameters) profiles along multi-span links is experimentally obtained via receiver-side (Rx) digital SI without any testing instruments such as OTDR. Excess losses (0, 2, and 5 dB) and different types of fibers—namely, standard single-mode fiber (SSMF), dispersion-shifted fiber (DSF), and nonzero DSF (NZ-DSF)—manually inserted in 280-km multi-span links can be successfully detected and distinguished. This work is an extension of the conference paper[13], and as additional works, we apply the method to long-haul systems (~2,080 km) and analyze various types of link conditions: the excess loss inserted, different fiber launch powers, and non-uniform level diagrams. The essential performance is evaluated in terms of the accuracy and measurement range of loss profiles with different fiber launch powers, which has not been reported in relevant works so far[7, 13]. Furthermore, we demonstrate that when the fiber launch power is varied (from -4 to 6 dBm) in certain spans (i.e. non-uniform level diagram), NN-based DBP correctly detects such deviated spans.

We also note that since the original paper[13] was published, several applications of the NN-based DBP method have been demonstrated, such as estimation of the multiple passband narrowing (PBN) at node optical filters[14] and estimation of multiple amplifier gain spectra[15]. The method will thus facilitate automated management of optical transmission systems because it can diagnose all fiber spans, multiple optical filters, and even inline amplifiers without dedicated equipment such as OTDR and optical spectrum analyzer. This will make it easier for anyone to build and use optical networks.

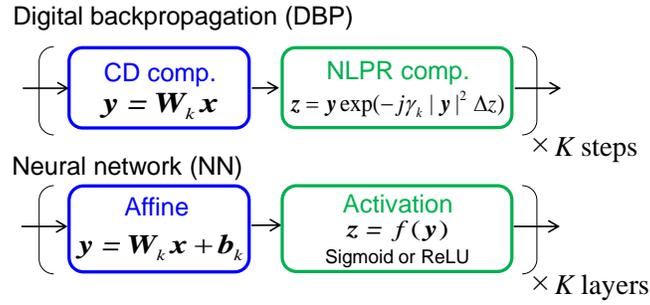

**Fig. 1 DBP and NN.** DBP has the same structure as an NN This similarity provides several advantages: (i) the physical interpretation of the DBP parameters is straightforward; (ii) better hyperparameter (nodes, layers, and activation function etc.) and initial values can be chosen due to their clear physical meaning; (ii) various optimization techniques developed in machine learning are available[9-12].

## Results

### NLSE and DBP

Our goal is to obtain the loss (or signal power) and dispersion profiles along optical fibers. For this purpose, we find the NLSE parameter distributions in actual transmission links from the boundary conditions, i.e., from the transmitted and received signals in the manner of SI. The NLSE describes how optical signals propagate in an optical fiber as follows:

$$\frac{\partial E'}{\partial z} = \left( -\frac{\alpha(z)}{2} - \frac{i}{2}\beta(z)\frac{\partial^2}{\partial t^2} \right) E' + i\gamma(z)|E'|^2 E'. \qquad (1)$$

Here, $E' = E'(z, t)$ is the envelope of the electric fields (optical signals) at position $z$ from the transmitter and time $t$. The boundary conditions are represented by $E(0, t)$ and $E(L, t)$, where $L$ denotes the total transmission length. $\alpha(z), \beta(z),$ and $\gamma(z)$ represent the loss (and amplification), group velocity dispersion, and nonlinear coefficient, respectively, and we assume that they are position-dependent. By substituting the normalized amplitude $E(z,t) = E'(z,t)\exp(\frac{1}{2}\int_0^z \alpha(z')dz')$, which has constant power during propagation, into Eq. (1), we can merge $\alpha(z)$ and $\gamma(z)$ as follows:

$$\frac{\partial E}{\partial z} = -\frac{i}{2}\beta(z)\frac{\partial^2}{\partial t^2}E + i\gamma'(z)|E|^2 E \equiv [\boldsymbol{D}(z) + \boldsymbol{N}(z)]E, \tag{2}$$

where $\gamma'(z) = \gamma(z)\exp(-\int_0^z \alpha(z')dz')$. In this study, $\gamma'(z)$ and $\beta(z)$ are the targets of estimation. If $\gamma(z)$ is constant, then $\gamma'(z)$ represents the accumulated loss and amplification and thus indirectly gives the signal power profile along a transmission link; meanwhile, $\beta(z)$ directly gives the dispersion distribution.

The NLSE is usually solved by a numerical analysis approach, SSFM[16]. In this algorithm, an approximated solution is obtained by assuming that the dispersion (first) term and the nonlinear (second) term operate independently when the signal propagates a short distance $\Delta z = z_{k+1} - z_k$ ($k=0,1,2\ldots K-1$), where $z_0=0$ and $z_K=L$. Specifically, iteration of the following equations gives the approximated optical signal $E(z,t)$ at an arbitrary position $z$:

$$B(z_{k+1},t) = \exp(\int_{z_k}^{z_{k+1}} \boldsymbol{N}(z')dz')E(z_k,t) \simeq \exp(\boldsymbol{N}(z_k)\Delta z)E(z_k,t), \tag{3}$$

$$E(z_{k+1},t) = \exp(\int_{z_k}^{z_{k+1}} \boldsymbol{D}(z')dz')B(z_{k+1},t) \simeq \exp(\boldsymbol{D}(z_k)\Delta z)B(z_{k+1},t), \tag{4}$$

where we use the rectangular approximation for the integrals. A more precise approximation such as the trapezoid rule can be used for higher accuracy. The exponential operator for dispersion can be deformed by using the Fourier operator $F$, giving $\exp(\boldsymbol{D}(z_k)\Delta z) = F^{-1}\exp(\frac{i}{2}\beta(z_k)\omega^2 \Delta z)F$. By applying this iterative operation backwards, the transmitted signal can be restored from the received digital signal $E(L, nT)$ ($n=0,1,2\ldots$) as follows:

$$\hat{E}(z_{k-1},nT) = \exp(-\boldsymbol{N}(z_k)\Delta z)\exp(-\boldsymbol{D}(z_k)\Delta z)\hat{E}(z_k,nT), \tag{5}$$

which constitutes a well-known fiber NLC technique, namely, digital backpropagation (DBP)[3]. Here, $T$ denotes the sampling period. As shown in Fig. 1, DBP entails concatenated iteration of linear chromatic dispersion compensation (CDC) and NLPR. In this work, we refer to a set of a linear operation and a nonlinear operation as a step. The iteration of a sufficient number of steps (a sufficiently small $\Delta z$) restores the transmitted signal $\hat{E}(0, nT)$ with high accuracy.

## Problem formulation and algorithm

In standard DBP, the NLSE parameters $\gamma'(z)$ and $\beta(z)$ that appear in each step are usually unknown. We find these parameters as optimal values that best emulate the propagated link, in which the received signals $E(L, nT)$ fully return to the transmitted signals $E(0, nT)$. This can be formulated as a common and classical problem of least squares in SI. The task is to find the optimal parameters as follows:

$$\hat{\beta}(z_k), \hat{\gamma}'(z_k) = \arg\min I = \arg\min \left\| \boldsymbol{E}(0, nT) - \hat{\boldsymbol{E}}(0, nT) \right\|^2. \tag{6}$$

This is a least-square problem for the parameters of the cascaded blocks (CDC and NLPR), and it can be solved by gradient descent using "error" backpropagation[17] as NN. Fig. 2 shows a block diagram of the algorithm we use, which consists of data preprocessing, NN-based DBP, and cost function $I$. Note that, in our error backpropagation, the partial derivatives required for gradient decent are calculated using the Wirtinger calculus because all the inputs and outputs of the DBP blocks are complex-valued.

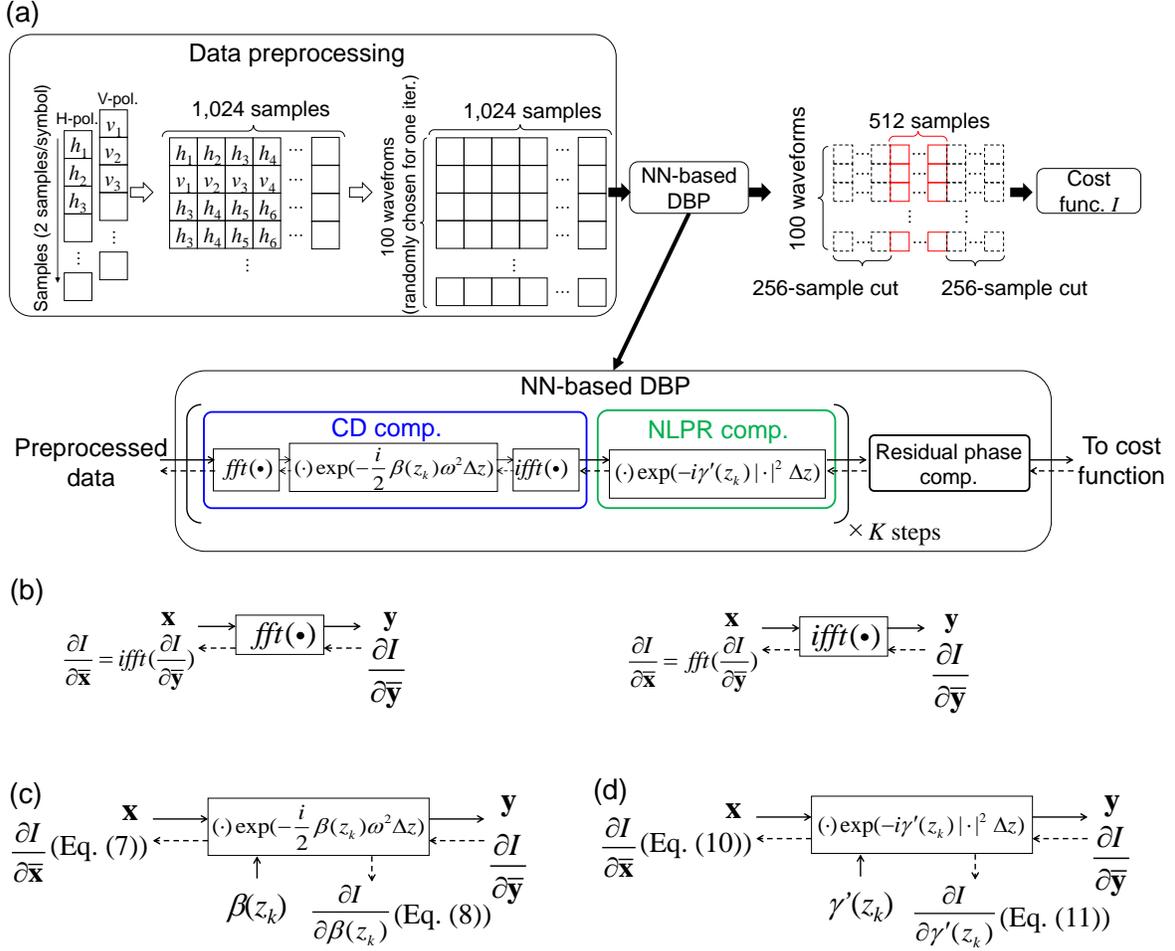

**Fig. 2 Algorithm for loss and dispersion profile extraction.** (a) Algorithm and data structures for NN-based DBP, whose estimation targets are $\gamma'(z_k)$ and $\beta(z_k)$. (c) Backpropagation of the FFT, which is equivalent to the IFFT of the backpropagated derivatives, and vice versa. (d) Backpropagation of the frequency-domain CDC. (d) Backpropagation of the NLPR.

As shown in Fig. 2(a), the input vectors are polarization demultiplexed signals, and for simplicity, the H- and V-polarization signals are treated as independent signals in a batch. We show an example of data size used for 280-km transmission, where 1,024 samples are taken out of each polarization every two samples and stored, and then 100 waveforms are randomly chosen from the stored data for a mini-batch.

An adjacent 1/2 overlap part is included in an input vector of a mini-batch to sufficiently compensate for the CD response and eliminated before the cost function *I*.

After data preprocessing, a mini-batch is fed into NN-based DBP, which consists of iteration of linear CDC and NLPR blocks, and residual phase rotation. In the *k*-th linear CDC block, linear operations are implemented with the fast Fourier transform (FFT), i.e., $\mathbf{y} = ifft\left[fft(\mathbf{x}) \cdot \exp(-\frac{i}{2}\beta(z_k)\omega^2 \Delta z)\right]$ to reduce the computational complexity and avoid keeping large matrices in random-access memory (RAM) for the backpropagation, where **x** and **y** denote the input and output of a block. As shown in Fig. 2(b), the backpropagation of the FFT is equivalent to the inverse FFT (IFFT) and vice versa. Also, the backpropagation of the frequency-domain CDC and the partial derivative for $\beta(z_k)$ update is calculated as (Fig. 2(c)):

$$\frac{\partial I}{\partial \overline{\mathbf{x}}} = \frac{\partial I}{\partial \mathbf{y}}\frac{\partial \mathbf{y}}{\partial \overline{\mathbf{x}}} + \frac{\partial I}{\partial \overline{\mathbf{y}}}\frac{\partial \overline{\mathbf{y}}}{\partial \overline{\mathbf{x}}} = 0 + \frac{\partial I}{\partial \overline{\mathbf{y}}}\exp(\frac{i}{2}\beta\omega^2 \Delta z), \tag{7}$$

$$\frac{\partial I}{\partial \beta} = \frac{\partial I}{\partial \mathbf{y}}\frac{\partial \mathbf{y}}{\partial \beta} + \frac{\partial I}{\partial \overline{\mathbf{y}}}\frac{\partial \overline{\mathbf{y}}}{\partial \beta} = 2\text{Re}\left(\frac{\partial I}{\partial \overline{\mathbf{y}}}\frac{\partial \overline{\mathbf{y}}}{\partial \beta}\right) = \text{Re}\left[(i\omega^2 \Delta z)\frac{\partial I}{\partial \overline{\mathbf{y}}}\overline{\mathbf{x}}\exp(\frac{i}{2}\beta\omega^2 \Delta z)\right], \tag{8}$$

where $\overline{\mathbf{x}}$ denotes complex conjugate of **x**. Thus, the backpropagation of the entire CDC block is simply implemented as:

$$\frac{\partial I}{\partial \overline{\mathbf{x}}} = ifft\left[fft(\frac{\partial I}{\partial \overline{\mathbf{y}}}) \cdot \exp(\frac{i}{2}\beta\omega^2 \Delta z)\right]. \tag{9}$$

Fig. 2(d) shows the *k*-th NLPR block, where the signal phase is nonlinearly derotated and its backpropagation is implemented as:

$$\frac{\partial I}{\partial \overline{\mathbf{x}}} = \frac{\partial I}{\partial \mathbf{y}}\frac{\partial \mathbf{y}}{\partial \overline{\mathbf{x}}} + \frac{\partial I}{\partial \overline{\mathbf{y}}}\frac{\partial \overline{\mathbf{y}}}{\partial \overline{\mathbf{x}}} = \frac{\partial I}{\partial \mathbf{y}}(-i\gamma'\Delta z)\mathbf{x}^2 \exp(-i\gamma'|\mathbf{x}|^2 \Delta z) + \frac{\partial I}{\partial \overline{\mathbf{y}}}(1 + i\gamma'|\mathbf{x}|^2 \Delta z)\exp(i\gamma'|\mathbf{x}|^2 \Delta z). \tag{10}$$

$$\frac{\partial I}{\partial \gamma'} = \frac{\partial I}{\partial \mathbf{y}}\frac{\partial \mathbf{y}}{\partial \gamma'} + \frac{\partial I}{\partial \overline{\mathbf{y}}}\frac{\partial \overline{\mathbf{y}}}{\partial \gamma'} = 2\text{Re}\left(\frac{\partial I}{\partial \overline{\mathbf{y}}}\frac{\partial \overline{\mathbf{y}}}{\partial \gamma'}\right) = \text{Re}\left[(2i\Delta z)\frac{\partial I}{\partial \overline{\mathbf{y}}}\overline{\mathbf{x}}|\mathbf{x}|^2 \exp(i\gamma'|\mathbf{x}|^2 \Delta z)\right], \tag{11}$$

After residual phase derotation, the MSE *I* is calculated via Eq. (6). Here, we can reconstruct the transmitted signals $E(0, nT)$ from the ordinary demodulation process in the Rx DSP, and thus, no pilot symbols or training sequences are needed. For better convergence, we use Adam[18] to update the NLSE parameters $\gamma'(z_k)$ and $\beta(z_k)$.

## Non-commutativity and related works

It has also been demonstrated that the NN-based DBP can even extract the responses of multiple optical filters separately with the insertion of complex FIR filters into the DBP blocks at positions corresponding to the optical filters[14]. As another technique to reveal the signal power profile, the correlation between the reference signal and the signals of which the NLPR is compensated at a position of interest, can be taken[7]. A notable commonality among these digital monitoring techniques is that they leverage the noncommutativity of linear and nonlinear operations to identify the position of physical characteristics. For example, in this study, we can estimate the pointwise dispersion $\beta(z_k)$ and nonlinear coefficients $\gamma'(z_k)$ because the (linear) CD and NLPR operators ( $\exp(\boldsymbol{D}(z_k)\Delta z)$ and $\exp(\boldsymbol{N}(z_k)\Delta z)$ ) are not commutative. For signals that have passed through multiple noncommutative systems, the inverse system that minimizes the cost function must be a reverse-order operation. Therefore, we can uniquely determine the order and positions of the systems by the minimum mean square criterion. In general, the same notion is also applicable to other noncommutative systems. In fact, the aforementioned study that separately estimated two optical filter responses used the fiber nonlinearity in between them, because a linear-nonlinear-linear (L-N-L) system has a unique inverse[14]. As another example, it has been demonstrated that the modulator driver nonlinearity in an optical transmitter can be effectively compensated by treating the transmitter structure—i.e., the digital-to-analog converter (DAC), drivers, and modulators—as a Wiener-Hammerstein model[19,20] (cascaded FIR filter, nonlinear function, and another FIR filter), which is an L-N-L system and thus noncommutative.

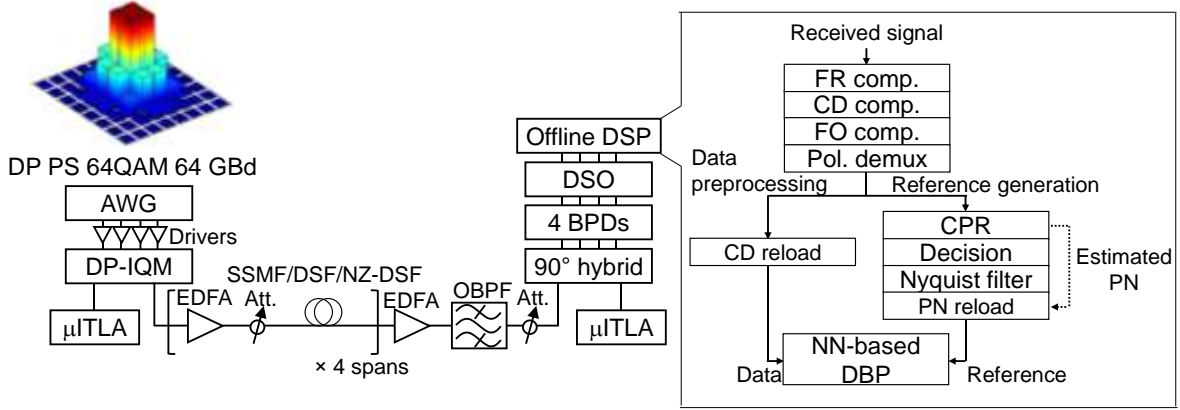

**Fig. 3 Experimental setup for straight-line 280-km transmission.** For loss profile acquisition, an attenuator is inserted in each span. For dispersion profile acquisition, the tested fibers were SSMF, DSF, and NZ-DSF. The results figures (4 and 5) depict the details of the transmission links. For NLSE parameter learning, the answer signal was reconstructed from the reference generation path in the Rx DSP. OBPF: optical bandpass filter; FR: frequency response; FO: frequency offset; PN: phase noise.

Loss and dispersion profile extraction for metro-reach links

We conducted experiments to examine the effectiveness of the NN-based DBP for a straight-line 4-span link. Fig. 3 shows the experimental setup and a block diagram of the offline DSP. A probabilistically shaped 64-QAM 64-GBd signal with a period of 65,536 symbols was generated with an information rate of 3.305 bits and an entropy of 4.347 bits, assuming a 21% forward error correction (FEC) overhead[21]. The signal was Nyquist-pulse-shaped by using a root-raised-cosine filter with a roll-off factor of 0.2; then the transmitter's frequency response (FR)[22] is compensated, followed by a resampling block for emission from a 120-GSa/s arbitrary waveform generator (AWG). The generated electrical signal was converted to an optical signal by 65-GHz drivers and a dual-polarization IQ modulator (DP-IQM). A micro-integrable tunable laser assembly (μITLA) with a linewidth of 40 kHz and a carrier wavelength of 1555.752 nm was used for the transmitter laser and the Rx local oscillator. The fiber input power was set to +5 dBm. The tested fibers were SSMF ($\alpha = 0.199$ dB/km, $D = 16.90$ ps/nm/km), DSF ($\alpha = 0.230$ dB/km, $D = 0.378$ ps/nm/km), and NZ-DSF ($\alpha$ = 0.225 dB/km, $D$ = 2.59 ps/nm/km). On the receiver side, the signal was post-amplified by an erbium-doped fiber amplifier (EDFA), filtered by a 5-nm optical bandpass filter (OBPF), and converted to electrical signals by a coherent receiver composed of a 90° hybrid and 100-GHz-bandwidth balanced photo detectors (BPDs). The received signals were digitized by a 160-GSa/s digital sampling oscilloscope (DSO) and demodulated offline in the Rx-DSP. There, the Rx FR was estimated beforehand and compensated[22]; this was followed by CDC. After frequency offset (FO) compensation[23], polarization demultiplexing was applied by an adaptive equalizer with a butterfly configuration. Then, the signal was divided into two paths: signal preprocessing to restore the dispersion and reshape the demodulated vectors, and transmitted signal reconstruction for gradient descent. The latter path consisted of carrier phase recovery (CPR)[24], symbol decision, the same Nyquist filtering as on the transmitter side, and reloading of the phase noise estimated in CPR. The length per DBP step (distance granularity) $\Delta z$ was a constant 2 km. The learning rate of Adam and the number of iterations for gradient descent were set to $1.0 \times 10^{-3}$ and 50, respectively. No regularization term was added to the cost function. The initial values of $\gamma'(z_k)$ and $\beta(z_k)$ were set to 0 and the average CD (i.e., *total CD / total DBP steps*), respectively. All the algorithms were coded in MATLAB, and the time to finish the learning process was within 1 minute on a single GPU 11 GB of memory. 20 profiles are averaged to enhance the signal to noise ratio (SNR).

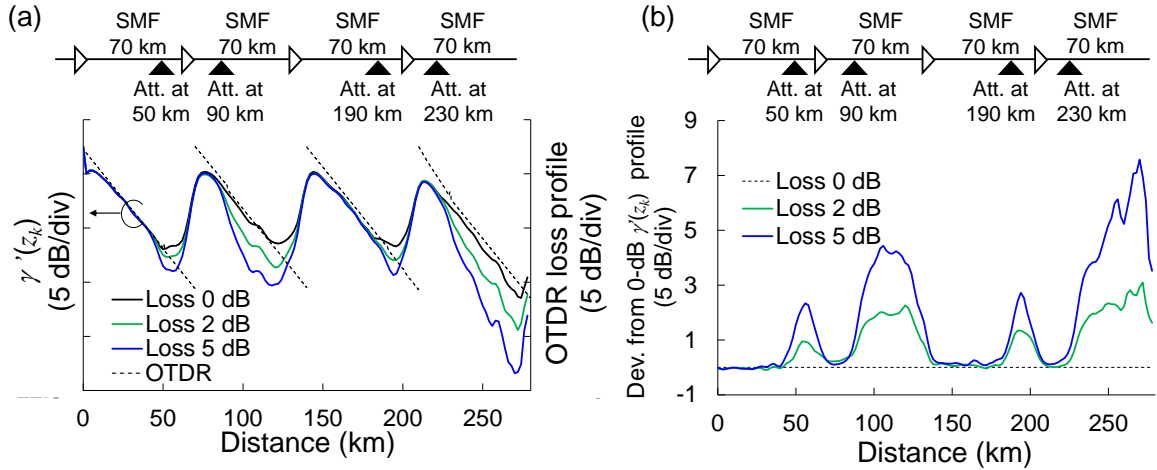

**Fig. 4 Results for loss (signal power) profile estimation.** (a) Obtained $\gamma'(z_k)$ distributions along four spans of SSMF. The OTDR loss profile is also shown for reference. The profiles reflect the fiber loss, amplification by the EDFA, and the inserted attenuators. (b) Differences between the normal (0-dB loss) and anomalous (2- and 5-dB losses) state profiles.

First, we obtained $\gamma'(z_k)$ (loss or signal power) profiles for the four spans of 70 km, as shown in Fig. 4(a). The OTDR power profile is also shown for reference. In all cases, the NLPR clearly reflected the fiber loss and the amplification by the EDFA. When 2-dB (green) and 5-dB (blue) attenuators were inserted at 50, 90, 190, and 230 km, reduction of $\gamma'(z_k)$ was observed due to the excess loss at the insertion points. Fig. 4(b) shows the difference between the normal (0-dB attenuation) and abnormal conditions. The beginnings of the peaks correspond to the anomaly loss points, and the relative attenuation level was successfully obtained. However, the peaks of the first and third span are lower than expected and the power profile fluctuated in the latter halves of the fourth spans. This was because the nonlinearity was too weak there as a result of the accumulated attenuation, which led to reduced measurement sensitivity.

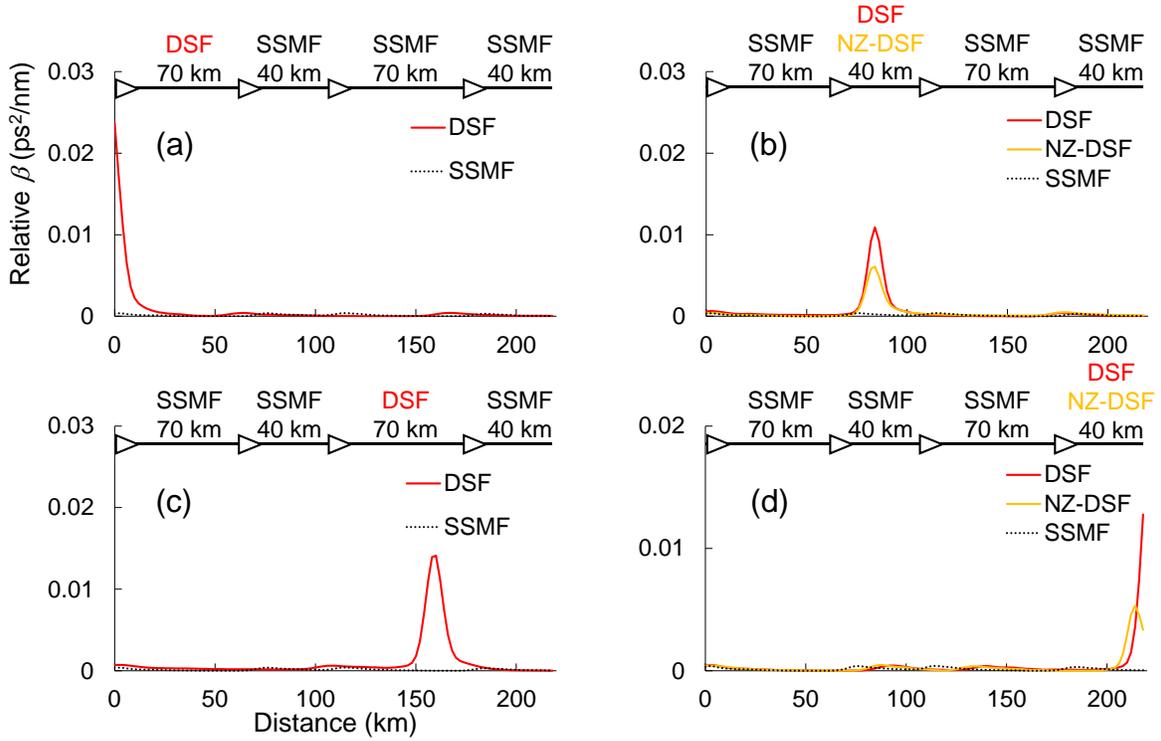

**Fig. 5 Results for dispersion profile estimation.** Obtained $\beta(z_k)$ profiles along four spans of SSMF with DSF and NZ-DSF spans mixed in. The DSF and NZ-DSF spans were inserted as the (a) first, (b) second, (c) third, and (d) fourth span of the link. The profiles clearly exhibit sharp peaks in the DSF spans.

Next, Fig. 5 shows the profiles obtained for the learned $\beta(z_k)$ in 4-span links with a DSF or NZ-DSF span inserted. The black dotted lines represent the SSMF-only links, and the solid lines represent the cases that included the DSF (red) and NZ-DSF (yellow) spans. The vertical axes indicate the relative $\beta(z_k)$ with respect to the minimum value over the whole link. Note that $\beta(z_k)$ was negative at the tested wavelength (1555.752 nm), and higher $\beta(z_k)$ (i.e., closer to zero) thus corresponds to lower dispersion. Compared to the SSMF-only cases, a sharp peak was observed when a DSF or NZ-DSF span was included, indicating the presence of a lower-dispersion fiber in the link. Also, the peaks with NZ-DSF were smaller than those with DSF, indicating that the NZ-DSF spans had higher dispersion. Thus, NN-based DBP could even distinguish the fiber types in a link. Ideally, however, the dispersion-shifted spans were expected to show not a sharp peak but a flat profile like a square wave over the whole span. The reason for the results is that the initial values of $\beta(z_k)$ were set to *total CD / K* and were thus already in a local minimum. Escaping from such a local minimum is quite hard, because once the total CD deviates from the optimal value, the signal quality (i.e., the cost function) greatly deteriorates, and the learned parameters diverge. Nevertheless, the profiles successfully detected the dispersion-shifted spans, and thus, we could take such measures as fiber input power adjustment or a DSF-span detour to maximize the transmission capacity.

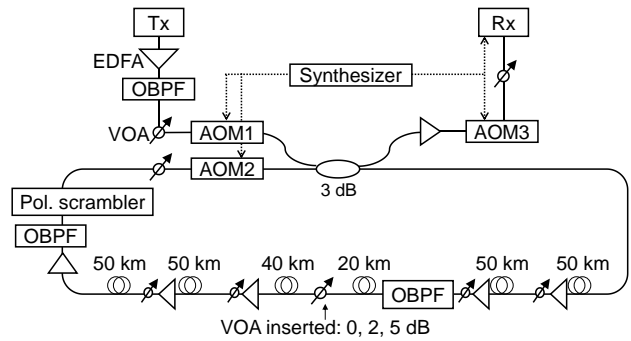

**Fig. 6 Experimental setup for circulating-loop transmission.** One circulation corresponded to five spans (260 km) of SSMF. A VOA was inserted 20 km from the beginning of the third span to emulate an anomaly loss (0, 2, and 5 dB). OBPF: optical bandpass filter; AOM: acousto-optic modulator.

Loss profile extraction for long-haul links and performance evaluation

To investigate the applicability of the NN-based DBP method to long-haul systems, we conducted circulating loop experiments in which the tested distance ranged from 260 (one circulation) to 2,080 km (eight circulations). Fig. 6 shows the setup. In this experiment, no DSF was inserted and the dispersion profile was not obtained, which we leave for a future work. The basic configuration of the transmitter and receiver was the same as in the previous section. Uniform 16-QAM signals were generated and then chopped with an acousto-optic modulator (AOM1) with a passing period of one circulation (260 km). The transmission line consisted of five spans of SSMF ($\alpha = 0.199$ dB/km, $D = 16.90$ ps/nm/km), and a variable optical attenuator (VOA) was inserted 20 km from the beginning of the third span to emulate an anomaly loss. The fiber launch power is varied by VOAs from -4 to 6 dBm. WaveShapers acted as OBPFs to cut the amplified spontaneous emission (ASE) noise and flatten the gain tilt. After a loop-synchronized polarization scrambler, AOM2 opened the gate for the duration of an arbitrary number of circulations and closed it while the signals were input to the circulating loop. AOM3 selected the desired circulated signals and cut instantaneous surges from the EDFA, and a 256-GSa/s DSO captured the signals and was synchronized to AOM3. For NN-based DBP, we assumed that the transmitted waveform was known as the answer signal for learning, because in the long-haul case, an input power of 6 dBm is too high to achieve the FEC limit due to the strong fiber nonlinearity. The mini-batch size was set to 20.

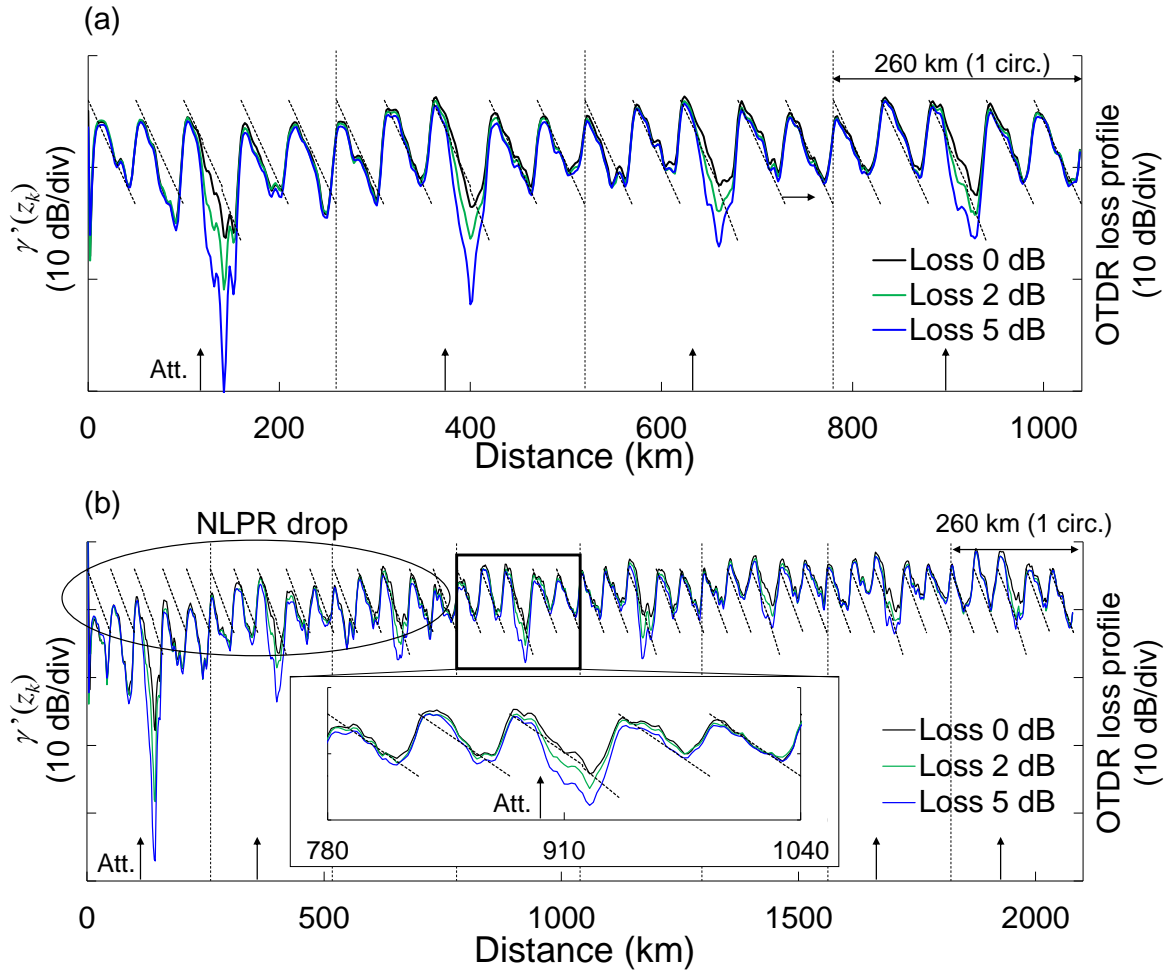

**Fig. 7 Loss profile results for long-haul transmission.** Obtained $\gamma'(z_k)$ profiles for transmission over (a) 1,040 km (four circulations) and (b) 2,080 km (eight circulations). The inset in (b) is an enlargement of the profile for the fourth circulation (from 780 to 1,040 km). In both (a) and (b), the profiles successfully detected the inserted loss; however, in the 2,080-km transmission, $\gamma'(z_k)$ dropped near the transmitter end and deviated from the expected profile.

Fig. 7(a) and (b) show the obtained loss profiles for transmission over 1,040 km (four circulations) and 2,080 km (eight circulations). The fiber input power was set to 6 dBm, and the attenuation by the VOA 20 km from the beginning of the third span was varied among 0, 2, and 5 dB. Note that tilt scaling was applied to the 0-dB loss profile to fit the OTDR reference, and the same scaling factor was used for the other profiles. One circulation corresponded to five spans (260 km), and the third span of each circulation clearly exhibited the loss due to the VOA, even in the 2,080-km transmission. Thus, the results support the applicability of NN-based DBP for long-haul system monitoring.

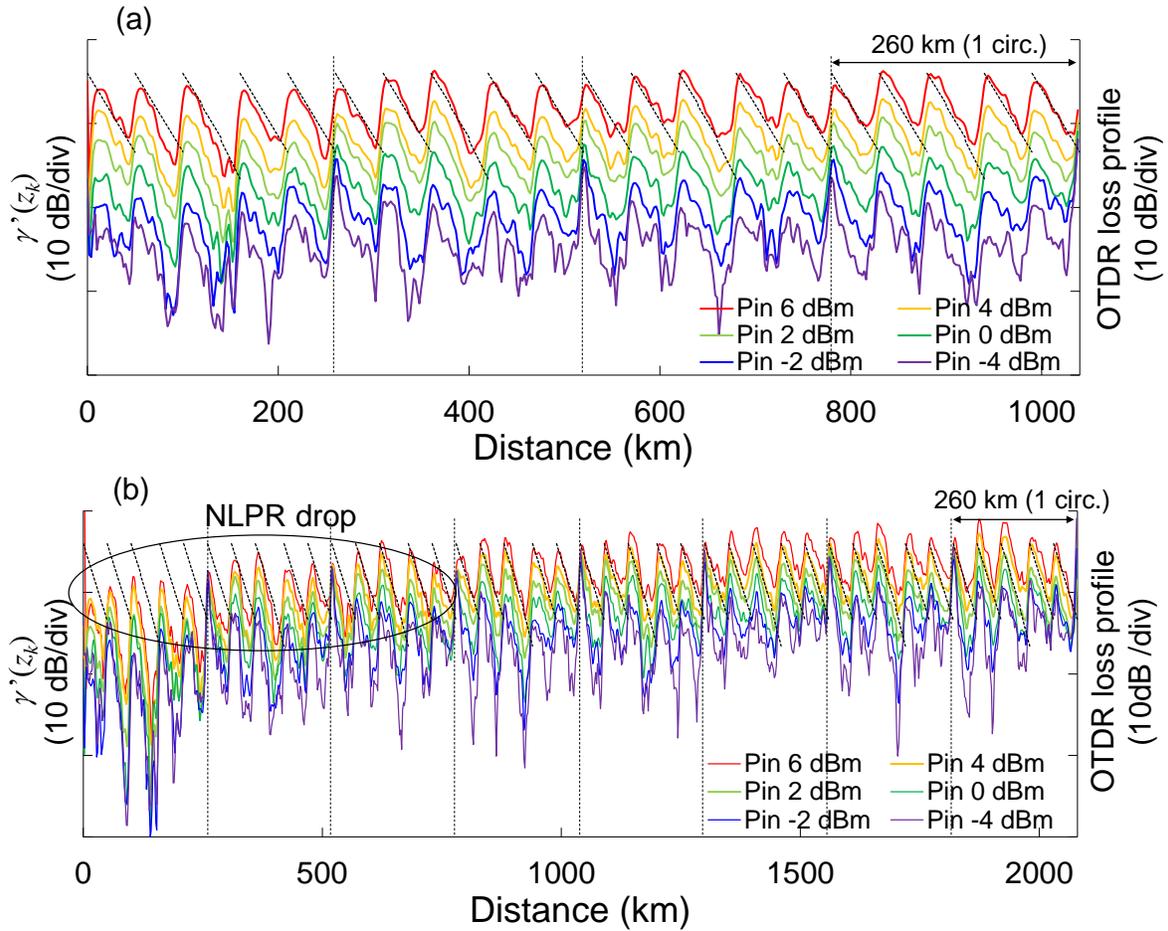

**Fig. 8 Input power dependency of the long-haul loss profiles.** Obtained $\gamma'(z_k)$ profiles for transmission over (a) 1,040 km (four circulations) and (b) 2,080 km (eight circulations) with different launch powers. The profiles deteriorated as the launch power decreased. As in Fig. 7, $\gamma'(z_k)$ tended to drop toward the transmitter end.

However, because NN-based DBP relies on the fiber nonlinearity for loss profile extraction, a higher fiber input power would be desirable to increase the accuracy; this is prohibitive in practical systems because increased fiber nonlinearity degrades the SNR. To determine the input power required to achieve a certain accuracy, we obtained power profiles for different input powers (from -4 to 6 dBm), as shown in Fig. 8. As in Fig. 7, the tilt of the 6-dBm profiles was scaled to fit the OTDR profile, and the same scaling factor was applied to the other profiles. Two observations are notable: (i) the $\gamma'(z_k)$ increased with the input power, which suggests that the estimated $\gamma'(z_k)$ was in good agreement with EDFA gains; and (ii) the profiles fluctuated significantly when the input power was low because the insufficient nonlinearity led to reduced sensitivity.

In the 2,080-km transmission, shown in Fig. 7(b) and Fig. 8(b), as the DBP steps approached the transmitter end, the estimated $\gamma'(z_k)$ became lower than it was expected to be. We can explain this drop with the help of a study by Fan et al.[12], where the a reasonable physical interpretation of the learned $\gamma'(z_k)$ in NN-based DBP is provided for one DBP step per span with a quasi-analytical approach; note, however, that the number of steps per span in our experiments was much larger (25 or 30). As signals backpropagate to the transmitter end with the cascaded DBP operations, the noises accompanying the signals, such as the ASE noise, the incompletely compensated Kerr nonlinear term, and even the Rx electrical noise, are enhanced. As a result, the learned $\gamma'(z_k)$ near the transmitter ($z_k \simeq 0$) tends to be small to avoid further noise enhancement, and thus, weak $\gamma'(z_k)$ is observed. Another reason for the reduced accuracy in the long-haul case is that the number of optimized parameters $\gamma'(z_k)$ is increased for a long transmission distance (here, $\Delta z$ was a constant 2 km), making parameters more likely to fall into local minima.

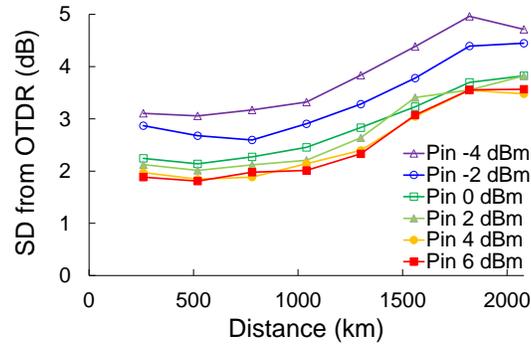

**Fig. 9 Accuracy as a function of distance.** Standard deviation (SD) from OTDR was degraded by the accumulated noise and the increased number of optimized parameters as the transmission distance increased.

To quantify the measurement range for a given accuracy, we plotted the standard deviation of the estimated profiles from the OTDR profile as a function of the transmission distance for different fiber input powers, as shown in Fig. 9. As we have already observed, the accuracy deteriorated when the fiber input power was low. Also, as the transmission distance increased, the deviation from the OTDR profile grew because of the increased accumulated noise and the increased number of parameters to be optimized. From this figure, we can determine the possible measurement range for a given accuracy and fiber input power. For example, suppose that we require a deviation of at most 2.5 dB from the OTDR profile. In that case, a fiber input power of -4 or -2 dBm is prohibitive for any transmission distance, but we can achieve a measurement range of 1,000 km with an input power of 0 dBm. If more input power is allowed (e.g., 4 dBm), then a 1,400-km measurement range is achievable. However, no significant difference was observed between the 4- and 6-dBm cases, which means that the accuracy saturates at a fiber input power of approximately 4-dBm, and further accuracy improvement cannot be expected.

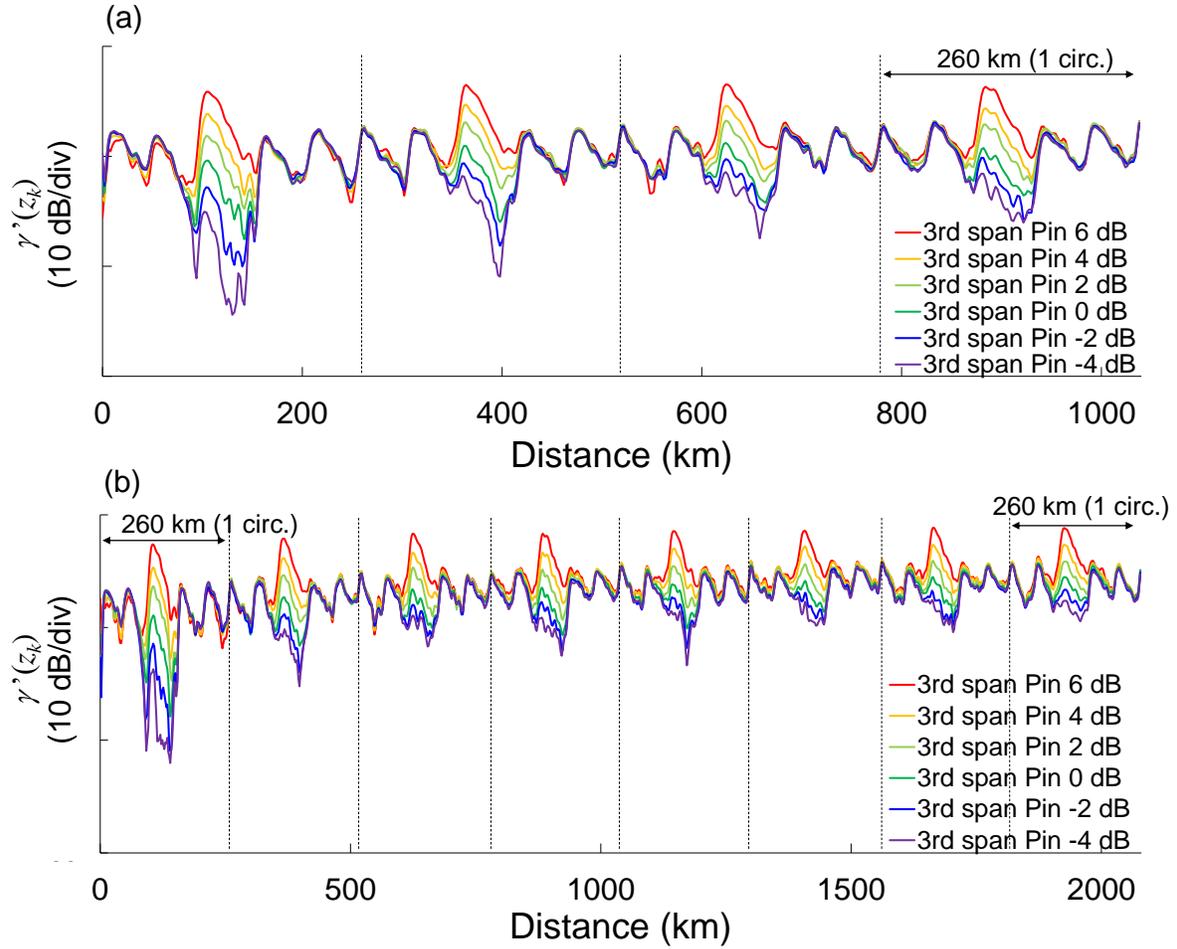

**Fig. 10 Optical level diagram extraction.** The launch power of the third span of the circulating loop was intentionally varied from -4 to 6 dBm, and that of the other spans was uniformly set to 2 dBm. The obtained $\gamma'(z_k)$ profiles correctly distinguish the modified spans, while the profiles for all the other spans are at the same level.

## Optical level diagram extraction

For the results discussed so far, we have assumed that the fiber input power is constant for all spans. If profiles obtained by NN-based DBP can automatically detect an anomalous input power for a span, then we can expect to implement functionality to remotely adjust the input power of such a span without direct measurement. Fig. 10 shows power profiles with different fiber input powers (-6 to 4 dBm) at the third span of the circulating loop. The input powers of the other spans were uniformly set to 2 dBm. The profiles clearly exhibit the non-uniform fiber input power at the third span of every five (one circulating loop), and the power at that span can be distinguished, while the power at the other spans is at the same level in all the profiles. This means that NN-based DBP can reveal the optical level diagram and the input power of each span from the Rx DSP.

Among potential applications of the obtained level diagrams, the amount of nonlinear noise can be estimated by feeding it to an analytical model, such as a Gaussian noise model[25,26] or the corresponding Python application, GNPy[27]. NN-based DBP could be especially useful for Raman-amplified systems, in which OTDR generally does not work because the backward Raman scattering acts as forward Raman scattering for the backscattered Rayleigh light. In forward Raman systems, the power peak appears in the middle of a span, which requires paying more attention to the nonlinear impact[28]. NN-based DBP can extract power profiles of such systems, thus facilitating the nonlinear management and maximizing the transmission capacity by adjusting the monitored power. Furthermore, by obtaining level diagrams of all wavelength-division multiplexed (WDM) channels, we can also unveil multiple amplifier gain spectra, because the level diagrams reveal the fiber launch powers of individual amplifiers. Accordingly, multiple Raman gain spectra for a multi-span Raman-amplified link have been successfully acquired separately by using NN-based DBP[15], which can also ease the management of high-power Raman amplifiers.

## Discussion

In this study, we have demonstrated physics-oriented learning of NLSE can work as a SI method and enable transmission line monitoring, revealing multi-span loss and dispersion profiles directly from the signal of interest. The results have shown that trivial physical characteristics such as anomaly losses and fiber input powers can be detected and localized. The method's applicability was investigated for not only metro-reach links but also long-haul systems, providing the tradeoff of accuracy vs. the measurement range, the input power dependency, and optical level diagrams. Machine learning techniques tend to focus on how close the output is to the answers, but by turning our attention to the optimized internal parameters of a model-based NN, we could keep a clear physical interpretation in those parameters and extract system information in detail. Note, however, that this work is only a proof of concept for the method as a monitoring technique, and there are many problems with it. Specifically, the required input power must be reduced in order to use actual data-carrying signals for this monitoring purpose, and the accuracy must be improved for practical, reliable use. In addition, in order for users to rely on this technique for performance prediction, we should thoroughly investigate its theoretical foundation, including the accuracy, spatial resolution, and measurement range. Nevertheless, the NN-based DBP method is attractive in that multi-span measurement can be done at once, and it has many potential use cases, as we have described. More importantly, the idea of this work can possibly be applied to SI for other PDE systems, such as the Navier-Stokes equations.

## Data Availability

The data files are available from the corresponding author upon reasonable request. Source data are provided with this paper.

## Code Availability

A portion of the code for training the DBP is available from the corresponding author upon reasonable request.

## Ethics declarations

The authors declare no conflict of interest.

## Supplementary Information

Source data are provided with this paper.